\documentclass[aps,prd,preprint,superscriptaddress,tightenlines,nofootinbib,showpacs]{revtex4}
\usepackage{graphicx}
\usepackage{dcolumn}
\usepackage{bm}
\usepackage{afterpage}
\usepackage{epsfig}

\newcommand{\etal}{{\it et al.}}

\begin{document}

\preprint{\tighten\vbox{\hbox{\hfil CLNS 06/1973}
                     \hbox{\hfil CLEO 06-16}
                       }}

\vspace*{10 mm}
\title
{Measurement of \boldmath{${\cal{B}}(\Upsilon(5S) \to
B_s^{(*)}\overline{B_s}^{(*)})$} Using \boldmath{$\phi$} Mesons}

\author{G.~S.~Huang}
\author{D.~H.~Miller}
\author{V.~Pavlunin}
\author{B.~Sanghi}
\author{I.~P.~J.~Shipsey}
\author{B.~Xin}
\affiliation{Purdue University, West Lafayette, Indiana 47907}
\author{G.~S.~Adams}
\author{M.~Anderson}
\author{J.~P.~Cummings}
\author{I.~Danko}
\author{J.~Napolitano}
\affiliation{Rensselaer Polytechnic Institute, Troy, New York 12180}
% ...
\collaboration{CLEO Collaboration} 
\noaffiliation

\date{October 12, 2006}

\begin{abstract}
 Knowledge of the $B_s$ decay fraction of the
 $\Upsilon$(5S) resonance, $f_S$, is important for $B_s$ meson studies at
 the $\Upsilon(5S)$ energy.  Using a data sample collected by the CLEO III detector at
 CESR consisting of 0.423 fb$^{-1}$ on the $\Upsilon$(5S) resonance,
 6.34 fb$^{-1}$ on the $\Upsilon$(4S) and 2.32 fb$^{-1}$ in the
 continuum below the $\Upsilon$(4S), we measure
 ${\cal{B}}(\Upsilon(5S)\to \phi X)=( 13.8\pm 0.7 ^{+2.3}_{-1.5}
 )\%$ and ${\cal{B}}(\Upsilon(4S)\to \phi X)=( 7.1 \pm 0.1 \pm 0.6
 )\%$; the ratio of the two rates is $(1.9 \pm 0.1 ^{+0.3}_{-0.2})$.
 This is the first measurement of the $\phi$ meson yield from the
 $\Upsilon$(5S). Using these rates,
and a model
 dependent estimate of ${\cal{B}}(B_s \to \phi X)$, we determine
  $f_S=(  24.6 \pm 2.9 ^{+11.0}_{-~5.3} )\%$. We also
 update our previous independent measurement of $f_S$ made
 using the inclusive $D_s$ yields to now be $(16.8 \pm 2.6 ^{+6.7}_{-3.4} )\%$, due to a
better estimate of the
 number of hadronic events. We also report the total
 $\Upsilon(5S)$ hadronic cross section above continuum to be
 $\sigma(e^+e^- \to \Upsilon(5S))=( 0.301 \pm 0.002 \pm 0.039)$
 nb. This allows us to extract the fraction of $B$ mesons as $(58.9\pm10.0\pm9.2)$\%, equal to 1-$f_S$.
 Averaging the three methods gives a model dependent result of $f_S=(21^{+6}_{-3})$\%.
\end{abstract}
 \maketitle


\section{Introduction}

The putative $\Upsilon$(5S) resonance was discovered at CESR long
ago by the CLEO~\cite{5s_cleo} and CUSB~\cite{5s_cusb}
collaborations by observing an enhancement in the total $e^+e^-$
annihilation cross-section into hadrons at a center-of-mass energy
of about 40 MeV above the $B_s^{*}\overline{B_s}^{*}$ production
threshold. Its mass and its production cross-section were measured
as (10.865$\pm$0.008)~GeV/$c^2$ and about 0.35~nb,
respectively~\cite{5s_cleo}.

The possible final states of the $\Upsilon$(5S) resonance decays
are: $B_s^*\overline{B_s}^*$, $B_s\overline{B_s}^*$,
$B_s\overline{B_s}$, $B^*\overline{B}^*$, $B\overline{B}^*$,
$B\overline{B}$, $B\overline{B}\pi$, $B\overline{B}^*\pi$,
$B^*\overline{B}^*\pi$, $B\overline{B}\pi\pi$, where the $B$ and
$\pi$ mesons can be either neutral or charged. Several models
involving coupled channel calculations have predicted the
cross-section and final state composition \cite{Models,UQM}. The
Unitarized Quark Model \cite{UQM}, for example, predicts that the
total $b \overline{b}$ cross-section at the $\Upsilon$(5S) energy is
dominated by $B^*\overline{B}^*$ and $B_{s}^*\overline{B}_{s}^*$
production, with $B_s$ production accounting for about 1/3 of the
total rate. The original $\sim$116 pb$^{-1}$ of data collected at
CESR did not reveal if any $B_s$ mesons were produced. By measuring
the inclusive $D_s$ production rate using CLEO III data we
previously measured the fraction of $B_s$ meson production at the
$\Upsilon$(5S), $f_S$, to be $(16.0 \pm 2.6 \pm 5.8)\%$ of the total
$b\overline{b}$ rate~\cite{Incl_Bs_5s}; this measurement uses a
theoretical estimate of ${\cal{B}}(B_s \to D_s X) = (92\pm
11)\%$~\cite{Incl_Bs_5s}.

$B_s$ production was confirmed in a second CLEO
analysis~\cite{Excl_Bs_5s} that fully reconstructed $B_s$ meson
decays. In addition, this analysis showed that the final states  are
dominated by the $B_s^*\overline{B}_s^*$ decay channel. These
results have been confirmed by the Belle collaboration~\cite{Belle}.
A third CLEO analysis of the exclusive $B$ reconstruction at the
$\Upsilon$(5S) showed that the $B\overline{B}X$ final states in
$\Upsilon$(5S) decays are dominated by $B^*\overline{B}^*$ with a
considerable contribution from $B\overline{B}^*$ and
$B^*\overline{B}$ final states~\cite{B_5s}.

Knowledge of the $B_s$ production rate at the $\Upsilon$(5S)
resonance, $f_S$, is necessary to compare with predictions of
theoretical models of $b$-hadron production. More importantly,
$f_S$ is also essential for evaluating the possibility of $B_s$
studies at the $\Upsilon$(5S) using current $B$-factories, and for
future $e^+ e^-$ Super-$B$ Factories, should they come to
fruition, where precision measurements of $B_s$ decays are an
important goal \cite{superb}. However, the determination of $f_S$
in a model independent manner requires several tens of
fb$^{-1}$~\cite{5s_Theory}. In this paper, we improve our
knowledge of the $B_s$ production at the $\Upsilon$(5S) by using
inclusive $\phi$ meson yields to make a second model dependent
measurement of ${\cal{B}}(\Upsilon(5S) \to
B_s^*\overline{B_s}^*)$.

We choose to examine $\phi$ meson yields because they will be
produced much more often in $B_s$ decays than in $B$ decays. The
branching fractions of $B_s$ to $D_s$ mesons and $B$ to $D$ mesons
are of order 1, while the rates of $B_s$ to $D$ mesons and $B$ to
$D_s$ mesons are on the order of 1/10. We also know that the
production rate of $\phi$ mesons is one order of magnitude higher in
$D_s$ decays than in $D^0$ and $D^+$ decays as measured using recent
CLEO-c data~\cite{D_Ds_incl}. These results are shown in
Table~\ref{tab:Dinc}. The rate of $D$ mesons into $\phi$ mesons is
only 1\%, while the rate of $D_s$ mesons into $\phi$ mesons is 16\%.

\begin{table}[ht]
\begin{center}
\caption{$D^0$, $D^+$ and $D_s^+$ inclusive branching ratios into
$\phi$ mesons. The $D$ meson decays used 281 pb$^{-1}$ of data at
the $\psi(3770)$, while the $D_s$ decays were measured using 195
pb$^{-1}$ at or near 4170 MeV.}
\begin{tabular}{lc}
\hline\hline &${\cal{B}}(D\to\phi X$){(\%)}\\\hline
 $D^0$  &  $1.05\pm 0.08\pm 0.07$\\
 $D^+$  &  $1.03\pm 0.10\pm 0.07$\\
 $D_s^+$&  $16.1\pm 1.2\pm 1.1$
\\\hline\hline
\end{tabular}
\label{tab:Dinc}
\end{center}
\end{table}

Since the $B\to \phi X$ branching ratio already has been measured to
be $(3.42\pm 0.13)\%$ \cite{PDG}, we expect a large difference
between the $\phi$ yields at the $\Upsilon$(5S) and at the
$\Upsilon$(4S), due to the presence of $B_s$, that we will use to
measure the size of the $B_s^{(*)}\overline{B_s}^{(*)}$ component at
the $\Upsilon$(5S). The analysis technique used here is similar to
the one used in \cite{Incl_Bs_5s}. We also present an update to the
first measurement of $f_S$ that used $D_s$ yields~\cite{Incl_Bs_5s},
and we report on the measurement of the total $\Upsilon$(5S)
hadronic cross-section. When we discuss the $\Upsilon$(5S) here, we
mean any production above what is expected from continuum production
of quarks lighter than the $b$ at an $e^+e^-$ center-of-mass energy
of 10.865 GeV.

The CLEO III detector is equipped to measure the momenta and
directions of charged particles, identify charged hadrons, detect
photons, and determine with good precision their directions and
energies. It has been described in detail previously
\cite{CLEO_DR,CLEO_RICH}.

\section{Measurement of \boldmath{${\cal{B}}(\Upsilon(5S) \to
B_s^{(*)}\overline{B_s}^{(*)})$} Using \boldmath{$\phi$} Meson
Yields}

\subsection{Data Sample and Signal Selection}

We use 6.34 fb$^{-1}$ integrated luminosity of data collected on
the $\Upsilon(4S$) resonance peak and 0.423 fb$^{-1}$ of data
collected on the $\Upsilon(5S)$ resonance ($E_{CM}=10.868~\rm
GeV$). A third data sample of 2.32 fb$^{-1}$ collected in the
continuum 40 MeV in center-of-mass energy below the $\Upsilon(4S)$
is used to subtract the four-flavor ($u$, $d$, $s$ and $c$ quark)
continuum events.

Hadronic events are selected using criteria based on the number of
charged tracks and the amount of energy deposited in the
electromagnetic calorimeter. To select ``spherical" $b$-quark
events we require that the Fox-Wolfram shape parameter \cite{Fox},
$R_2$, be less than $0.25$. $\phi$ meson candidates are looked for
through the reconstruction of a pair of oppositely charged tracks
identified as kaons. These tracks are required to originate from
the main interaction point and have a minimum of half of the
maximal number of hits in the tracking chambers. They also must
satisfy kaon identification criteria that uses information from
both the Ring Imaging Cherenkov (RICH) and the ionization loss in
the drift chamber, $dE/dx$, of the CLEO III detector. Kaon
identification has been described in detail previously
\cite{Incl_Bs_5s}.

\subsection{\boldmath{$\phi$} Meson Yields From  \boldmath{$\Upsilon$}(5S) and
\boldmath{$\Upsilon$}(4S) Decays}

\begin{figure}[htbp]
\centerline{\epsfig{figure=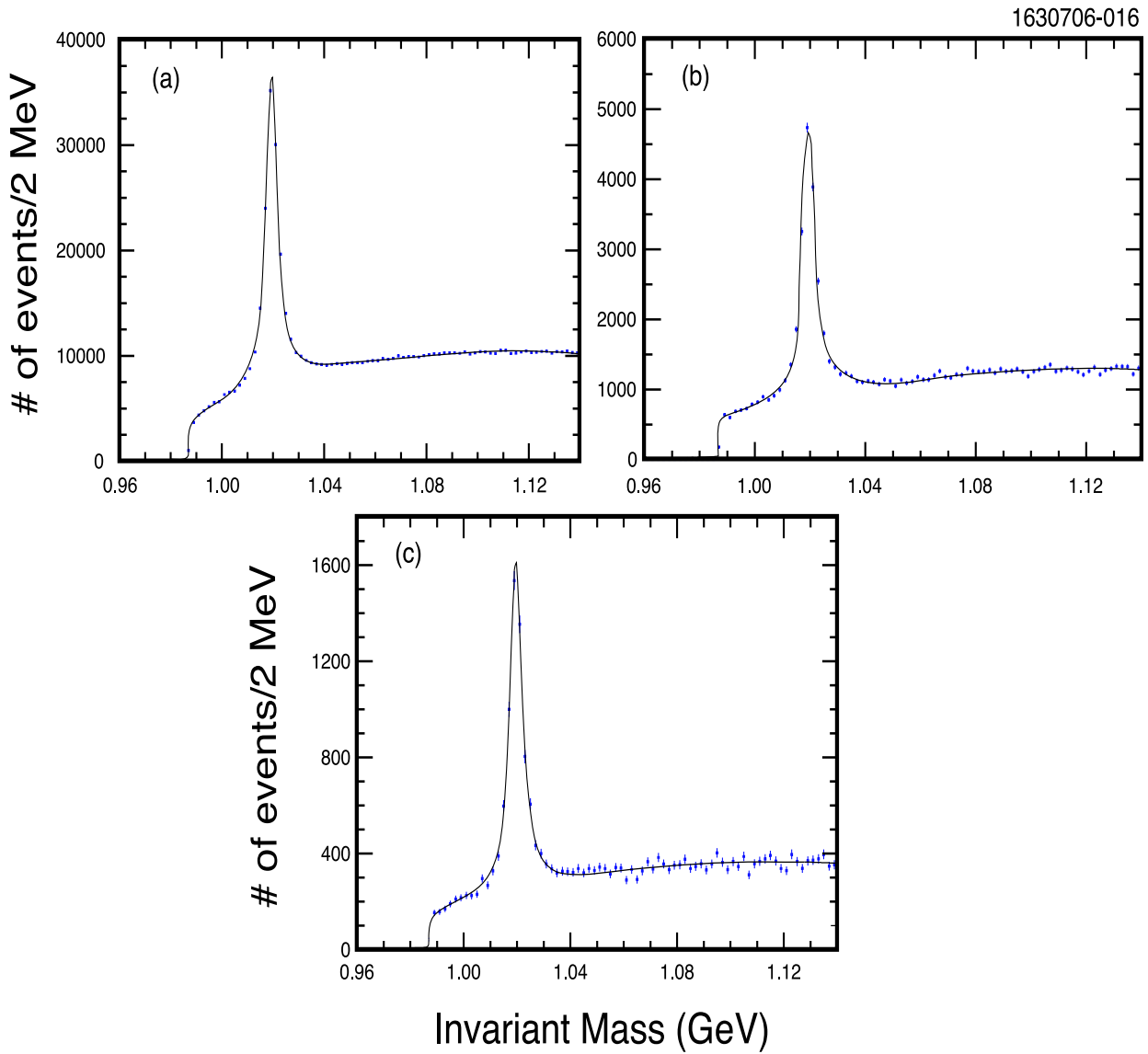,height=6.0in,width=6.5in}}
  \caption{\label{4s-5s-phi-inv-mass} The invariant mass distributions of the $\phi$ candidates
  with $x\leq$ 0.5 from: (a) the $\Upsilon$(4S) on-resonance
  data (b) the continuum below the $\Upsilon$(4S)
resonance data, and (c) the $\Upsilon$(5S) on-resonance data. The
solid line is the fit to the signal and background shapes explained
in the text.}
\end{figure}

All pairs of oppositely charged kaon candidates were examined for
$\phi$ candidates if their summed momenta is less than half of the
beam energy. Instead of momentum we choose to work with the variable
$x$ which is the $\phi$ momentum divided by the beam energy, to
remove differences caused by the the change of energies between
continuum data taken just below the $\Upsilon$(4S), at the
$\Upsilon$(4S) and at the $\Upsilon$(5S). The $K^+K^-$ invariant
mass distributions for $x \leq 0.5$ are shown in
Fig.~\ref{4s-5s-phi-inv-mass}.

\subsubsection{\boldmath{$K^+K^-$} Invariant Mass Spectra and Yields}

A crucial aspect in this analysis is to accurately model the signal
and background shapes. For the signal we use a Breit-Wigner signal
shape convoluted with a Gaussian. The width of the Breit-Wigner
function was fixed to the natural width of the $\phi$ meson,
$\Gamma_1=4.26~\rm MeV$~\cite{PDG}. It is convoluted here with a
Gaussian function to allow the integration of the detector
resolution into the signal function. A second Gaussian is added for
an adequate fitting of the tails. The form of the signal fitting
function, where the dependent variable $\zeta$ is the invariant mass
of our $K^+K^-$ candidates, is given by
\begin{equation}
S(\zeta)=h(\zeta)+\int_{\zeta-5\sigma}^{\zeta+5\sigma}{f(y)g(\zeta-y)dy}.
\end{equation}
Here, at every physical point $\zeta$, the Breit-Wigner function
\begin{equation}
f(y)= \frac{1}{2\pi} \cdot \frac{A_1\cdot{\Gamma_1}}{[
(y-\overline{y})^2 + \frac{\Gamma_1^2}{4} ]}~
\end{equation}
is convoluted with the Gaussian
\begin{equation}
g(\zeta-y)= \frac{1}{\sqrt{2\pi}\sigma_2}~\exp\left[
\frac{(\zeta-y)^2}{2\sigma_2^2}\right]~.
\end{equation}
The integration over $y$ is between the limits $\zeta-5\sigma$, and
$\zeta+5\sigma$. $A_1$ and $\overline{y}$ are the area and mean
value of the Breit-Wigner, and $\sigma_2$ is the standard deviation
of the Gaussian. $h(\zeta)$ is a second Gaussian added to fit the
tails.

The background shape is given by a function, chosen to model the
threshold, that has the following functional form:
\begin{equation}
B(\zeta)=A\cdot{{(\zeta-\zeta_0)}^p}\cdot{\exp~{[c_1(\zeta-\zeta_0)
+ c_2(\zeta-\zeta_0)^2]}},
\end{equation}
where $A$ is the normalization, $\zeta_0$ is the threshold, $p$ is
the power of a polynomial about the turn-on point, and $c_1$ and
$c_2$ are linear and quadratic coefficients in the exponential.
Because this function has a sharp rise followed by an exponential
tail, we are able to accurately describe the threshold behavior at
low invariant mass close to the kinematic limit.

\begin{figure}[htbp]
 \centerline{\epsfig{figure=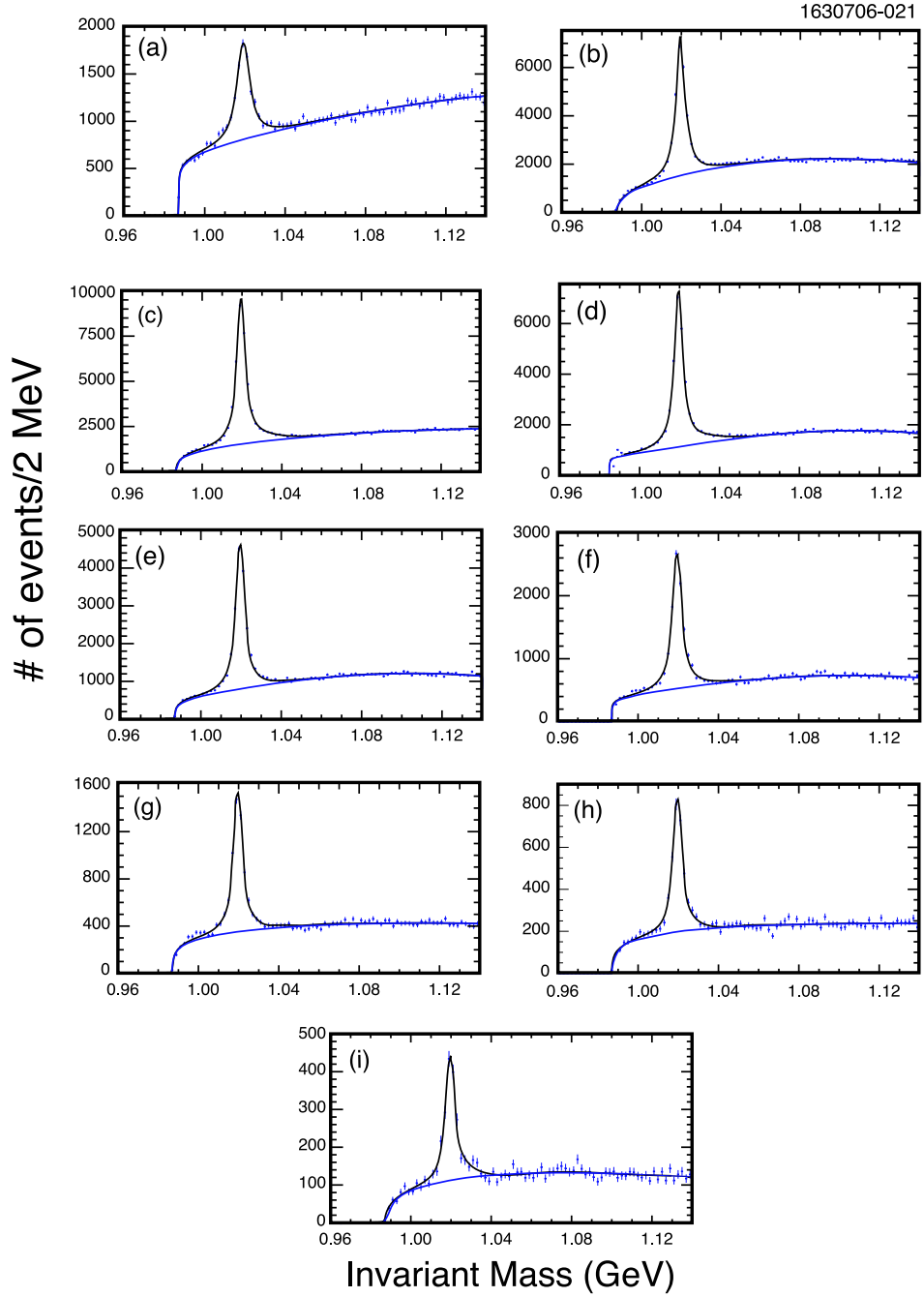,height=7in}}
  \caption{\label{PhiMass4sOnx} The $K^+K^-$ mass combinations from $\Upsilon$(4S) on-resonance data, fitted to
  the sum of a Breit-Wigner signal shape centered at the nominal $\phi$ mass, convoluted with a
  Gaussian to describe the detector resolution, and a second Gaussian distribution for a better
  parametrization of the tails. The background is parameterized by
  a threshold function (see text). These distributions are in the $x$ intervals:
  (a) $0.05<x<0.10$,
  (b) $0.10<x<0.15$, (c) $0.15<x<0.20$, (d) $0.20<x<0.25$, (e) $0.25<x<0.30$,
  (f) $0.30<x<0.35$, (g) $0.35<x<0.40$, (h) $0.40<x<0.45$, (i) $0.45<x<0.50$.}
   \end{figure}

\begin{figure}[htbp]
 \centerline{\epsfig{figure=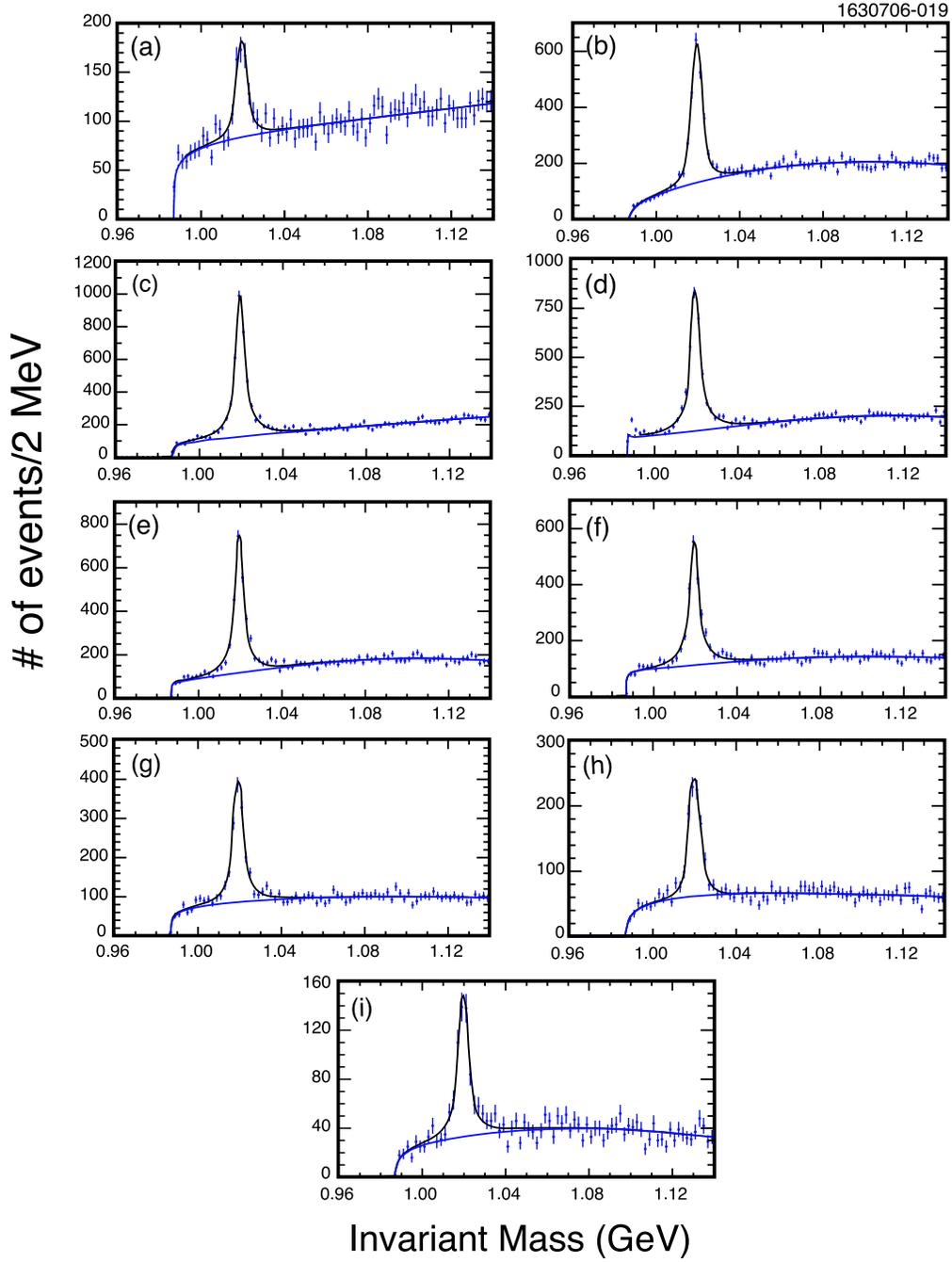,height=7in}}
  \caption{\label{PhiMass4sOffx} The $K^+K^-$ mass combinations from the
  continuum below the $\Upsilon$(4S), fitted to the sum of a
  Breit-Wigner signal shape centered at the nominal $\phi$ mass, convoluted with a
  Gaussian to describe the detector resolution, and a second Gaussian distribution for a better
  parametrization of the tails. The background is parameterized by
  a threshold function (see text). These distributions are in the $x$ intervals: (a) $0.05<x<0.10$,
  (b) $0.10<x<0.15$, (c) $0.15<x<0.20$, (d) $0.20<x<0.25$, (e) $0.25<x<0.30$,
  (f) $0.30<x<0.35$, (g) $0.35<x<0.40$, (h) $0.40<x<0.45$, (i) $0.45<x<0.50$.}
   \end{figure}

\begin{figure}[htbp]
 \centerline{\epsfig{figure=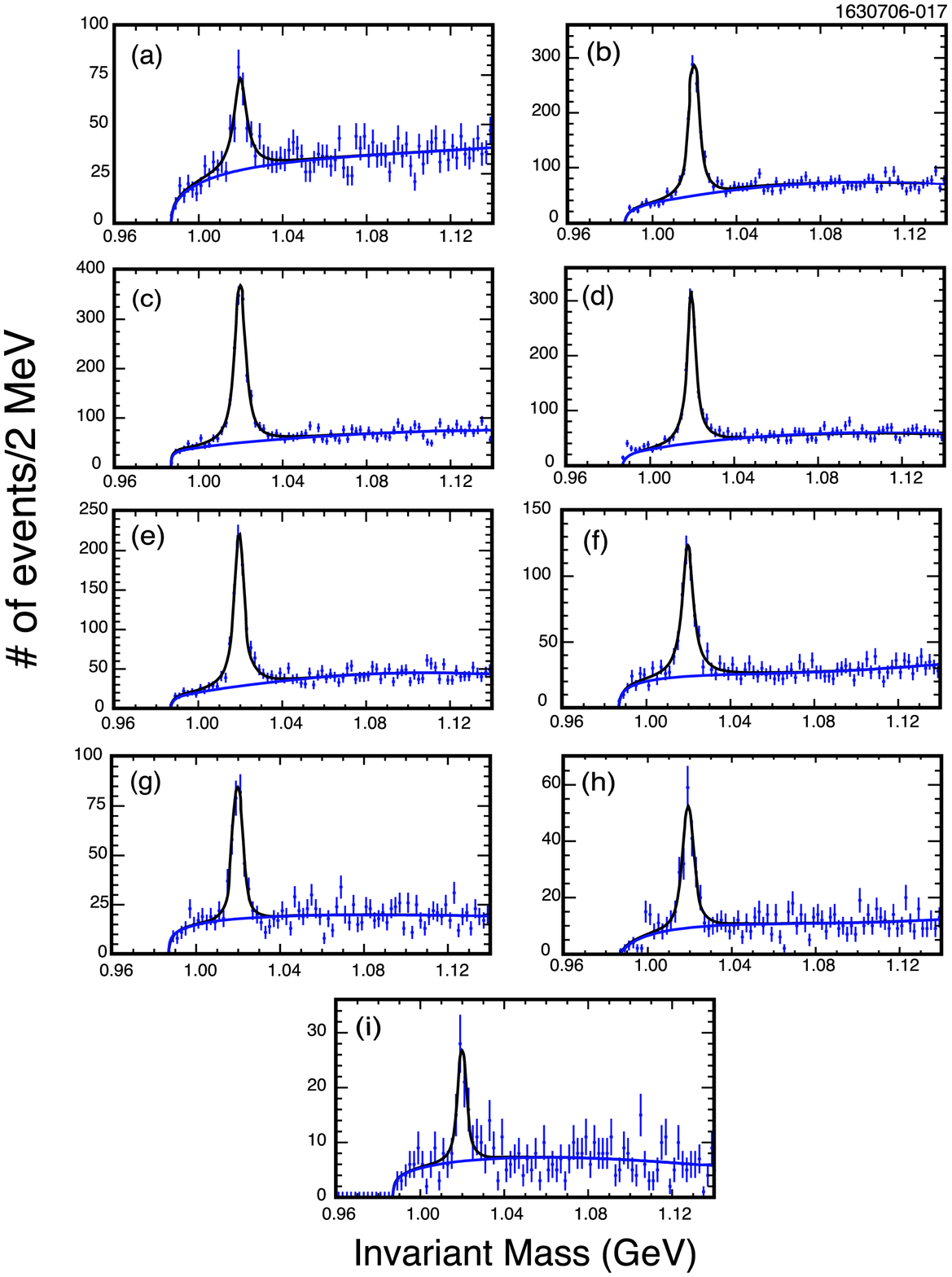,height=7in}}
  \caption{\label{PhiMass5sx} The $K^+K^-$ mass combinations from the $\Upsilon$(5S), fitted to the sum of a
  Breit-Wigner signal shape centered at the nominal $\phi$ mass, convoluted with a
  Gaussian to describe the detector resolution, and a second Gaussian distribution for a better
  parametrization of the tails. The background is parameterized by
  a threshold function (see text). These distributions are in the $x$ intervals:
  (a) $0.05<x<0.10$,
  (b) $0.10<x<0.15$, (c) $0.15<x<0.20$, (d) $0.20<x<0.25$, (e) $0.25<x<0.30$,
  (f) $0.30<x<0.35$, (g) $0.35<x<0.40$, (h) $0.40<x<0.45$, (i) $0.45<x<0.50$.}
   \end{figure}

\begin{figure}[htbp]
 \centerline{\epsfig{figure=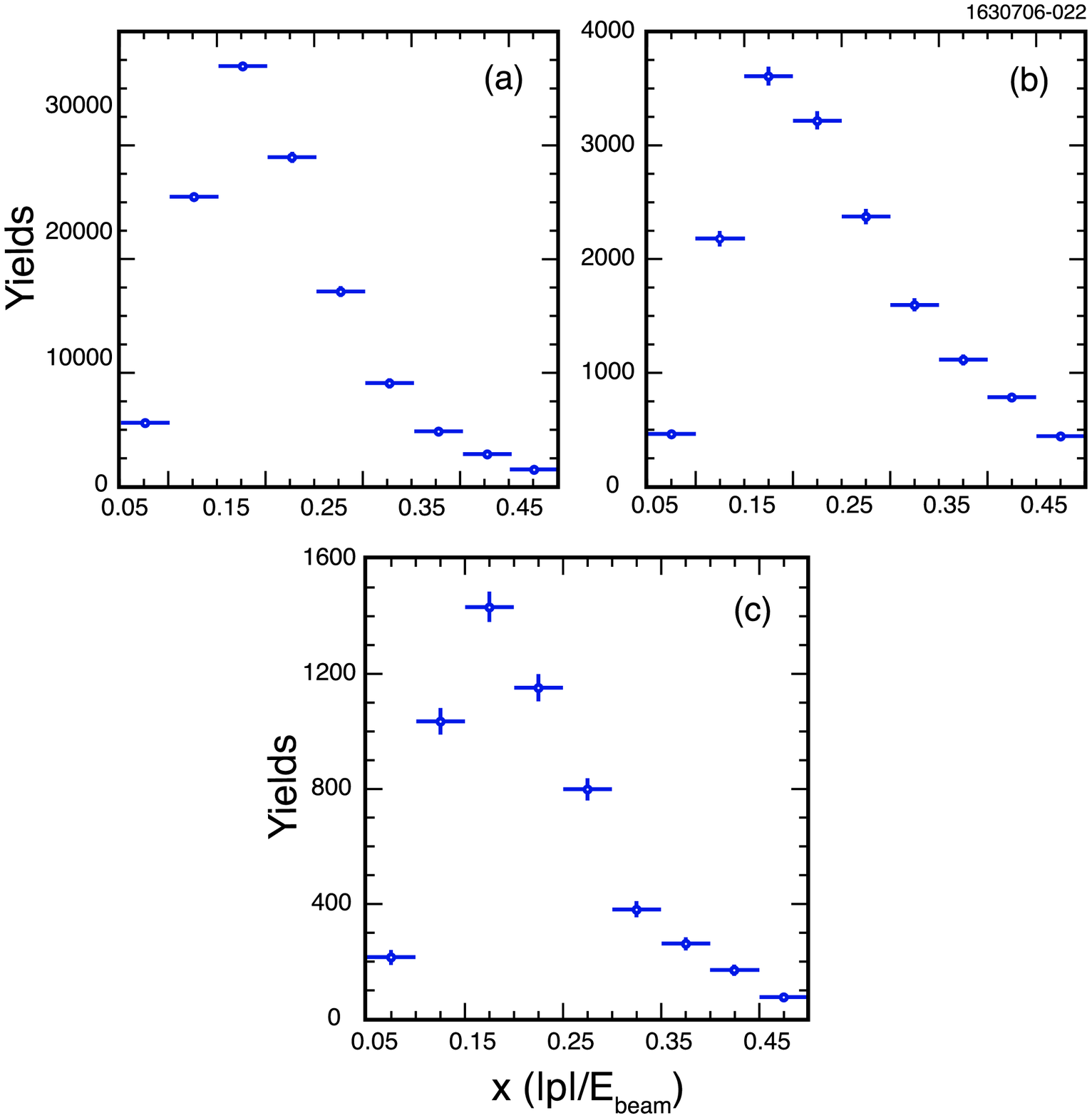,height=7.0in,width=6.8in}}
  \caption{\label{RawYields} Raw $\phi$ yields from:
  (a) the $\Upsilon$(4S) data, (b) the continuum below the $\Upsilon$(4S)
  data, (c) the $\Upsilon$(5S) data.}
\end{figure}

We show the invariant mass of the $K^+K^-$ candidates in 9 different
$x$ intervals (from 0.05 to 0.50) for all the data samples in
Fig.~\ref{PhiMass4sOnx}, Fig.~\ref{PhiMass4sOffx} and
Fig.~\ref{PhiMass5sx}.
We determine the raw yield of $\phi$'s shown in Fig.~\ref{RawYields}
(a), (b) and (c) from fits of the invariant mass distributions to
the functions $S(\zeta)$ and $B(\zeta)$ defined above. All the
parameters in $S(\zeta)$ in all the fits, except the yields, were
fixed to the values obtained when fitting the corresponding
distributions from the sum of the three data samples collected at
the three different beam energies in each $x$ interval separately.
The parameters describing $B(\zeta)$, on the other hand, are allowed
to float except for $\zeta_0$ which is fixed to twice the $K^+$
mass. The yields are listed in the second and third columns of
Table~\ref{tab:PhiRecon4S} and Table~\ref{tab:PhiRecon5S}.

The systematic errors on these yields are estimated to be $\pm 4\%$,
and are obtained by varying the fitting techniques. One variation is
to float the fitting parameters at the different beam energies,
instead of fixing them. We also used different functions including a
Breit-Wigner signal shape with and without convolution, or a double
Guassian shape for the signal, and for background either other
threshold functions or polynomials.

\begin{table}[htbp]
\begin{center}
\caption{$\phi$ yields from the $\Upsilon$(4S)
($N^i_{\Upsilon(4S)on}$), from the continuum below the
$\Upsilon$(4S) ($N^i_{cont}$), and from the continuum subtracted
$\Upsilon$(4S) ($N^{i}_{\Upsilon(4S)}$). Also listed are the $\phi$
reconstruction efficiencies ($\epsilon^{i}$), and the partial
$\Upsilon(4S)\to \phi X$ branching fractions as a function of $x$.
The systematic errors on the yields are $\pm 4\%$.}
\label{tab:PhiRecon4S}
\begin{tabular}{cccccc}
\hline\hline
 $x^{i}$($\frac{|p^{i}|}{E_{beam}}$)~ &~ $N^i_{\Upsilon(4S)on}$ &
$N^i_{cont}$ &
$N^{i}_{\Upsilon(4S)}$  & $\epsilon^{i}(\%)$ & $B^{i}$(\%)\\
 \hline
0.05-0.10 & $ 4938.1\pm135.2$ &~ $  462.4\pm41.0 $ &~ $3683.5 \pm175.1$ & $8.0 $  & $1.46\pm0.09$\\
0.10-0.15 & $22848.9\pm228.7$ &~ $ 2179.0\pm69.0 $ &~ $16937.3\pm295.5$ & $31.9$  & $1.68\pm0.04$\\
0.15-0.20 & $33207.4\pm269.1$ &~ $ 3608.7\pm84.1 $ &~ $23417.2\pm352.8$ & $47.1$  & $1.57\pm0.03$\\
0.20-0.25 & $26020.4\pm237.6$ &~ $ 3218.7\pm80.7 $ &~ $17288.3\pm323.1$ & $51.3$  & $1.06\pm0.03$\\
0.25-0.30 & $15364.1\pm178.8$ &~ $ 2374.2\pm69.4 $ &~ $ 8923.0\pm259.7$ & $49.1$  & $0.58\pm0.02$\\
0.30-0.35 & $8107.8\pm128.2 $ &~ $ 1597.6\pm57.8 $ &~ $ 3773.5\pm202.5$ & $42.3$  & $0.28\pm0.02$\\
0.35-0.40 & $4263.5\pm95.9  $ &~ $ 1113.4\pm48.6 $ &~ $ 1242.9\pm162.9$ & $35.8$  & $0.11\pm0.01$\\
0.40-0.45 & $2475.3\pm75.1  $ &~ $  786.6\pm42.3 $ &~ $  341.2\pm137.1$ & $21.6$  & $0.05\pm0.02$\\
0.45-0.50 & $1256.9\pm57.6  $ &~ $  443.6\pm30.9 $ &~ $   53.5\pm101.8$ & $13.4$  & $0.01\pm0.02$\\
\hline\hline
\end{tabular}
\end{center}
\end{table}

\begin{table}[htbp]
\begin{center}
\caption{$\phi$ yields from the $\Upsilon$(5S)
($N^i_{\Upsilon(5S)on}$), from the continuum below the
$\Upsilon$(4S) ($N^i_{cont}$ are the same as in the previous table),
and from the continuum subtracted $\Upsilon$(5S) continuum
subtracted ($N^{i}_{\Upsilon(5S)}$). Also listed are the $\phi$
reconstruction efficiencies ($\epsilon^{i}$ which are taken to be
the same as in the previous table), and the partial $\Upsilon(5S)
\to \phi X$ branching fractions as a function of $x$. The systematic
errors on the yields are $\pm 4\%$.} \label{tab:PhiRecon5S}
\begin{tabular}{cccccc}
\hline\hline
   $x^{i}$($\frac{|p^{i}|}{E_{beam}}$) & $N^i_{\Upsilon(5S)on}$ &
$N^i_{cont}$ & $N^{i}_{\Upsilon(5S)}$  &  $\epsilon^{i}(\%)$  &  $B^{i}$(\%)  \\
 \hline
0.05-0.10 & $ 214.7\pm26.2$ &~ $  462.4\pm41.0 $ &~ $135.3\pm27.1$ & $8.0 $  & $2.7 \pm0.6$\\
0.10-0.15 & $1034.8\pm46.0$ &~ $ 2179.0\pm69.0 $ &~ $660.4\pm47.5$ & $31.9$  & $3.3 \pm0.2$\\
0.15-0.20 & $1432.5\pm53.0$ &~ $ 3608.7\pm84.1 $ &~ $812.4\pm54.9$ & $47.1$  & $2.8 \pm0.2$\\
0.20-0.25 & $1151.9\pm47.3$ &~ $ 3218.7\pm80.7 $ &~ $598.9\pm49.3$ & $51.3$  & $1.9 \pm0.2$\\
0.25-0.30 & $ 789.2\pm38.6$ &~ $ 2374.2\pm69.4 $ &~ $390.2\pm40.4$ & $49.1$  & $1.3 \pm0.1$\\
0.30-0.35 & $ 382.3\pm27.9$ &~ $ 1597.6\pm57.8 $ &~ $107.8\pm29.7$ & $42.3$  & $0.4 \pm0.1$\\
0.35-0.40 & $ 261.2\pm22.7$ &~ $ 1113.4\pm48.6 $ &~ $ 70.0\pm24.1$ & $35.8$  & $0.3 \pm0.1$\\
0.40-0.45 & $ 169.7\pm19.1$ &~ $  786.6\pm42.3 $ &~ $ 34.5\pm20.4$ & $21.6$  & $0.3 \pm0.2$\\
0.45-0.50 & $  75.0\pm13.6$ &~ $  443.6\pm30.9 $ &~ $ -1.2\pm14.6$ & $13.4$  & $-0.1\pm0.2$\\
\hline\hline
\end{tabular}
\end{center}
\end{table}

\subsubsection{Continuum Subtraction}
\label{sec:cont_subtr} The numbers of hadronic events and $\phi$
candidates from the $\Upsilon$(5S) and the $\Upsilon$(4S) resonance
decays are determined by subtracting the scaled four-flavor ($u$,
$d$, $s$ and $c$ quarks) continuum events from the $\Upsilon$(4S)
and the $\Upsilon$(5S) data. The scale factors, $S_{nS}$, are
determined using the same technique described in \cite{Incl_Bs_5s},
where
\begin{equation}
S_{nS}= {\frac{L_{nS}}{L_{\rm cont}}}\cdot{ \left({\frac{E_{\rm
cont}}{E_{nS}}}\right)^2}~\label{eq:S_formula}
\end{equation}
with $L_{nS}$, $L_{\rm cont}$, $E_{nS}$ and $E_{\rm cont}$ being the
collected luminosities and the center-of-mass energies at the
$\Upsilon$(nS) and at the continuum below the $\Upsilon$(4S). We
find:

\begin{equation}
S_{4S}={2.713 \pm 0.001 \pm 0.027 }~\label{eq:4s_scale_Factor1}
,\end{equation} and
\begin{equation}
S_{5S}={(17.18 \pm 0.01 \pm 0.17
)}\cdot{10^{-2}}~.\label{eq:5s_scale_Factor1}
\end{equation}

We estimate the systematic error on these scale factors by obtaining
them in a different manner. Here we measure the ratio of the number
of charged tracks at the different beam energies in the $0.6<x<0.8$
interval. The lower limit on the $x$ interval used here is
determined by the maximum value tracks from $B\overline{B}$ events
can have, including smearing because of the measuring resolution,
and the upper limit is chosen to eliminate radiative electromagnetic
processes. Since the tracks should be produced only from continuum
events, we suppress beam-gas and beam-wall interactions, photon pair
and $\tau$ pair events using strict cuts on track multiplicities,
event energies and event shapes. Since particle production may be
larger at the higher $\Upsilon (5S)$ energy than the continuum below
the $\Upsilon (4S)$, we apply a small multiplicative correction of
(1.016$\pm$0.011)\%, as determined by Monte Carlo simulation to the
relative track yields. Track counting gives a 1\% lower value for
$S_{5S}$ and we use this difference as our estimate of the
systematic error. Note, that the error on the beam energy (0.1\%)
has a negligible effect.

These numbers are updated from those reported in \cite{Incl_Bs_5s},
due to an increase in Monte Carlo statistics, and the use of a more
precise calibration of the beam energy.

\subsubsection{\boldmath{$\phi$} Reconstruction Efficiency}

\begin{figure}[htbp]
\centerline{\epsfig{figure=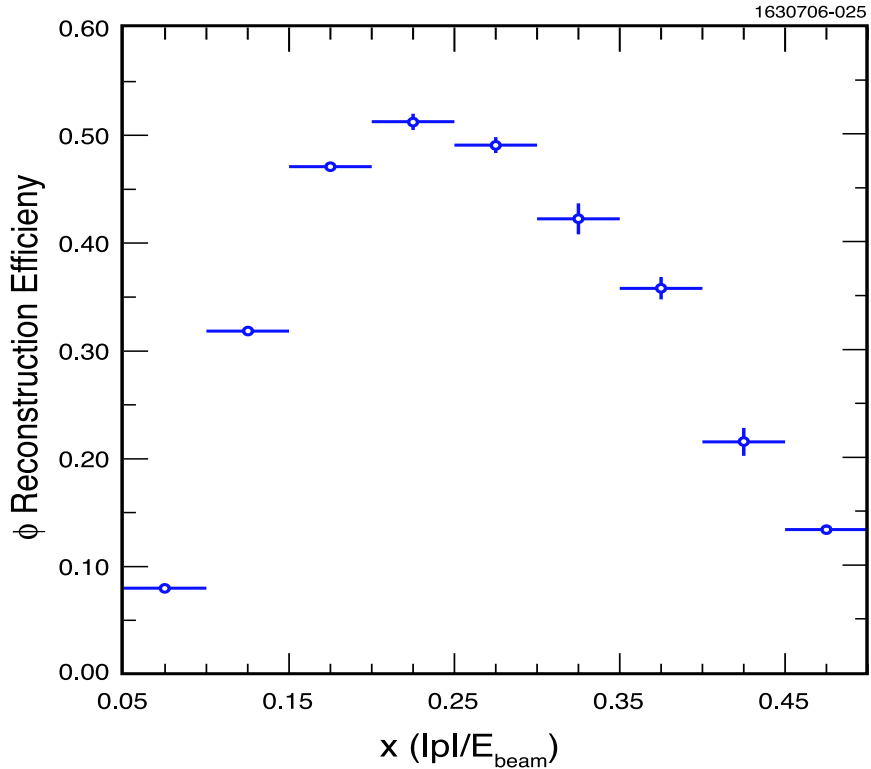,height=4in,width=4.5in}}
  \caption{\label{efficiency} $\phi$ reconstruction efficiency from more than 8 million $B\overline{B}$
  Monte Carlo simulated events. The errors are statistical only.}
\end{figure}

The $x$ dependent $\phi$ detection efficiency shown in
Fig.~\ref{efficiency} is determined by reconstructing and fitting
$\phi$ candidates in more than 8 million simulated $B\overline{B}$
generic decays.
The low reconstruction efficiency at small $x$ is due to the fact
that as the $\phi$ becomes less energetic, it becomes more probable
that it is formed of a slow kaon (with momentum below $0.2~\rm
GeV/c$) whose detection is very inefficient, since it is likely to
be absorbed in the beam-pipe or vertex detector material. The
efficiency for large $x$ ($0.4<x<0.5$) is somewhat lower than
naively expected because of the $R_2$ cut of 0.25. Due to the low
efficiency and large backgrounds in the first bin, $x<0.05$, we do
not measure the $\phi$ yield in this interval, but will rely on a
model of $\phi$ production to extract a value.

\subsubsection{\boldmath{$\phi$} Branching Ratios}

To find the number of hadronic decays produced above the four-flavor
continuum, denoted as $N^{Res}_{\Upsilon(nS)}$, we multiply the
number of events found in the continuum below the $\Upsilon$(4S),
$N^{off}_{\Upsilon(4S)}$, by the $S_{4S}$ and $S_{5S}$ scale
factors, and subtract them from the number of hadronic events found
at each resonance, $N^{on}_{\Upsilon(nS)}$:

\begin{equation}
N^{\rm Res}_{\Upsilon(4S)}=N^{\rm on}_{\Upsilon(4S)}-S_{4S}\cdot
N^{\rm off}_{\Upsilon(4S)}=(6.42\pm 0.01\pm
0.26)\cdot{10^{6}}~\label{eq:4s_had_events}
\end{equation}
\begin{equation}
N^{\rm Res}_{\Upsilon(5S)}=N^{\rm on}_{\Upsilon(5S)}-S_{5S}\cdot
N^{\rm off}_{\Upsilon(5S)}=(0.127\pm 0.001 \pm
0.016)\cdot{10^{6}}~.\label{eq:5s_had_events}
\end{equation}

Using these numbers together with $N^{i}_{\Upsilon(4S)}$
($N^{i}_{\Upsilon(5S)}$) and $\epsilon^{i}$ which are the $\phi$
yield and $\phi$ reconstruction efficiency in the $i$-th $x$
interval respectively, we measure the inclusive partial decay rate
of the $\Upsilon$(nS) resonance into $\phi$ mesons in the $i$-th
$x$ interval as follows:
\begin{equation}
{\cal{B}}^{i}(\Upsilon(nS)\to \phi
X)={\frac{1}{N^{Res}_{\Upsilon(nS)} \times {{\cal{B}}(\phi\to
K^+K^-)}}}~\left({\frac{N^{i}_{\Upsilon(nS)}}
{\epsilon^{i}}}\right)~\label{eq:RSoln}\\
\end{equation}

\begin{figure}[htb]
 \centerline{\epsfig{figure=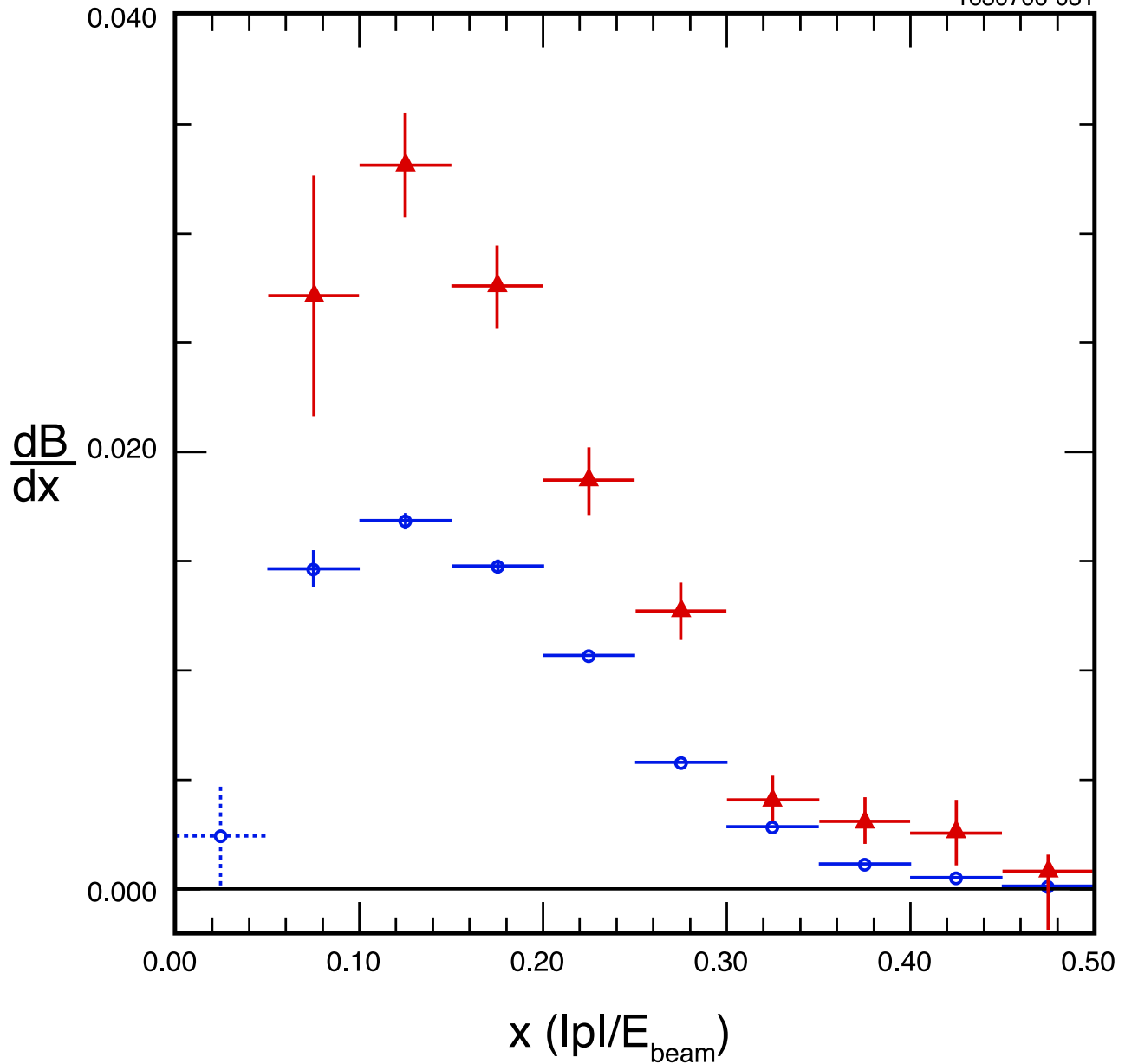,height=6.5in}}
  \caption{\label{PhiBrs} The efficiency corrected $\phi$ branching ratio in each $x$ interval
   from the $\Upsilon$(5S) resonance decays (triangles),
  and from the $\Upsilon$(4S) resonance decays (open circles).
  Errors are statistical only, except for the first bin where the $\Upsilon$(4S) data point
is a theoretical estimate based on Monte Carlo, with an error equal
to its value. }
\end{figure}

The results are listed in the last column of Table
\ref{tab:PhiRecon4S} and Table \ref{tab:PhiRecon5S} and shown in
Fig.~\ref{PhiBrs}.
Summing the results for $0.05<x<0.50$ we measure
\begin{equation}
{\cal{B}}^{(x>0.05)}(\Upsilon(4S)\to \phi X)=( 6.82 \pm 0.09 \pm
0.55)\%.~\label{4s-to-phi-9}
\end{equation}

\afterpage{\clearpage}

 As the rate in the first interval
0.00$<x<$0.05 is experimentally not accessible in this analysis, we
estimate the branching fraction in this interval using our Monte
Carlo simulation to model $\phi$ production. This interval has 3.6\%
of the total $\phi$ yield, which corresponds to a branching fraction
of 0.24\%. We take the error as equal to the value. The total
production rate then is

\begin{equation}
{\cal{B}}(\Upsilon(4S)\to \phi X)=( 7.06 \pm 0.09 \pm 0.60 )
\%,~\label{4s-to-phi}
\end{equation}
that when divided by two gives the $B$ meson branching fraction into
$\phi$ mesons of
\begin{equation}
{\cal{B}}(B\to \phi X)=( 3.53 \pm 0.05 \pm 0.30
)\%~,\label{eq:Btophiss}
\end{equation}
which is in good agreement with the PDG \cite{PDG} value of $(3.42
\pm 0.13)\%$. For the $\Upsilon$(5S) we measure for $x>0.05$
\begin{equation}
{\cal{B}}^{(x>0.05)}(\Upsilon(5S)\to \phi X)=( 12.9 \pm 0.7
^{+2.2}_{-1.4} )\%.~\label{5s-to-phi}
\end{equation}

In addition, we find for $x>0.05$

\begin{equation}
\frac{{\cal{B}}(\Upsilon(5S)\to \phi X)}{{\cal{B}}(\Upsilon(4S)\to
\phi X)}= 1.9 \pm 0.1 ^{+0.3}_{-0.2}~.\label{eq:ratio-to-phi}
\end{equation}

This result is more than $9.6\sigma$ significant, including both the
statistical and the systematic errors, thus demonstrating an almost
factor of two larger production of $\phi$ mesons at the
$\Upsilon$(5S) than at the $\Upsilon$(4S).

\subsubsection{Determination of The Statistical and Systematic
Uncertainties} \label{sec:stat_syst_errors}


Since the measurements presented in the previous section depend on a
large number of parameters, the corresponding errors on those
measurements can have highly correlated contributions. In this
section, we explain the method we used to extract the statistical
and the systematic errors on our measurements by taking into account
the sources of correlations between the different parameters.

 We measure the $\phi$
production rate at the $\Upsilon$(4S) and $\Upsilon$(5S) using
\begin{eqnarray}
\lefteqn{{{\cal{B}}^{(x>0.05)}(\Upsilon({\rm nS})\to \phi X)}
=\frac{1}{N^{\rm Res}_{\Upsilon(nS)}\cdot{\cal{B}}_1}
~{{\sum_{i=2}^{10}{\frac{N^{i}_{\Upsilon(nS)}}
{\epsilon^{i}}}}}}\nonumber\\
& & ~~~~~~~~~~~~~~~~=\frac{1}{({N^{\rm h}_{\Upsilon(nS)on}-S_n \cdot
N^h_{\rm cont}})\cdot{\cal{B}}_1}~
{{\sum_{i=2}^{10}{\frac{N^{i}_{\Upsilon(nS)on}-S_n \cdot N^{i}_{\rm
cont}} {\epsilon^{i}}}}}~.~\label{eq:syst_4SToPhi}
\end{eqnarray}

The quantities in this equation are defined as:

\begin{itemize}

\item
$N^{Res}_{\Upsilon(nS)}$ is the number of hadronic events at the
$\Upsilon$(nS) energy after continuum subtraction.

\item
$N^{\rm h}_{\Upsilon(nS)on}$ and $N^h_{\rm cont}$ are respectively
the number of hadronic events from the data taken at the
$\Upsilon$(nS) energy and from the continuum data taken at $30~\rm
MeV$ below the $\Upsilon$(4S) resonance.

\item ${\cal{B}}_1$ is the ${\cal{B}}(\phi \to K^+K^-)$ branching
fraction of 49.1\% \cite{PDG}.

\item
$N^{i}_{\Upsilon(nS)}$ is the number of $\phi$ candidates in the
$i$-th $x$ interval from the $\Upsilon$(nS) resonance events (i.e.
after continuum subtraction).

\item
$N^{i}_{\Upsilon(nS)on}$ and $N^{i}_{\rm cont}$ are the number of
$\phi$ candidates in the $i$-th $x$ interval from the
$\Upsilon$(nS)-ON resonance data and from the data taken at the
continuum below the $\Upsilon$(4S) respectively.

\item
$\epsilon^i$ is the reconstruction efficiency of $\phi$ candidates
in the $i$-th $x$ interval.

\item
$S_n$ are the continuum subtraction scale factors described in the
previous sections.

\end{itemize}

The ratio of the $\phi$ production rate from the $\Upsilon$(5S) to
the $\phi$ production rate from the $\Upsilon$(4S) is given by

\begin{eqnarray}
\lefteqn{ R={{{\cal{B}}^{(x>0.05)}(\Upsilon({\rm 5S})\to \phi X)}
\over {{\cal{B}}^{(x>0.05)}(\Upsilon({\rm 4S})\to \phi X)}} =
 {\frac {~~
\frac{1}{N^{Res}_{\Upsilon(5S)}\cdot{\cal{B}}_1}
~{{\sum_{i=2}^{10}{\frac{N^{i}_{\Upsilon(5S)}} {\epsilon^{i}}}}}~}
{~\frac{1}{N^{Res}_{\Upsilon(4S)}\cdot{\cal{B}}_1}
~{{\sum_{i=2}^{10}{\frac{N^{i}_{\Upsilon(4S)}} {\epsilon^{i}}}}}~}}
}\\\nonumber & & ~~~~~~~~~~~~~~~~~~~~~~~~~= {\frac{~
\frac{1}{({N^h_{\Upsilon(5S)on}-S_5*N^h_{\rm
cont}})\cdot{\cal{B}}_1}~
{{\sum_{i=2}^{10}{\frac{N^{i}_{\Upsilon(5S)on}-S_5*N^{i}_{\rm cont}}
{\epsilon^{i}}}}}~} {~\frac{1}{({N^h_{\Upsilon(4S)on}-S_4*N^h_{\rm
cont}})\cdot{\cal{B}}_1}~
{{\sum_{i=2}^{10}{\frac{N^{i}_{\Upsilon(4S)on}-S_4*N^{i}_{\rm cont}}
{\epsilon^{i}}}}}~}} ~.~\label{eq:syst_Ratio}
\end{eqnarray}
The independent parameters are: $S_4$ and $S_5$, $\epsilon^i$,
$N^{i}_{\Upsilon(nS)on}$ and $N^{i}_{\rm cont}$.  The error on
$\epsilon^i$ includes, as shown in Table~\ref{tab:eff}, a $2\%$
error on the tracking efficiency, a $2\%$ error on the particle
identification, both per track. In addition, it has a contribution
of $1\%$ from the error on the fitting method. The total systematic
error on the $\phi$ reconstruction efficiency is therefore equal to
$5.7 \%$, and is correlated among all the bins.
\begin{table}[htb]
\begin{center}
\caption{Systematic errors on the $\phi$ detection efficiency.}
\label{tab:eff}
\begin{tabular}{lc}
\hline\hline \multicolumn{2}{c}{Systematic error (\%)} \\ \hline
Track finding (per track)& 2 \\
Particle identification (per track) & 2 \\
Fitting techniques & 1\\
\hline
Total & 5.7 \\
 \hline\hline
\end{tabular}
\end{center}

\end{table}

\begin{figure}[htbp]
\centerline{\epsfig{figure=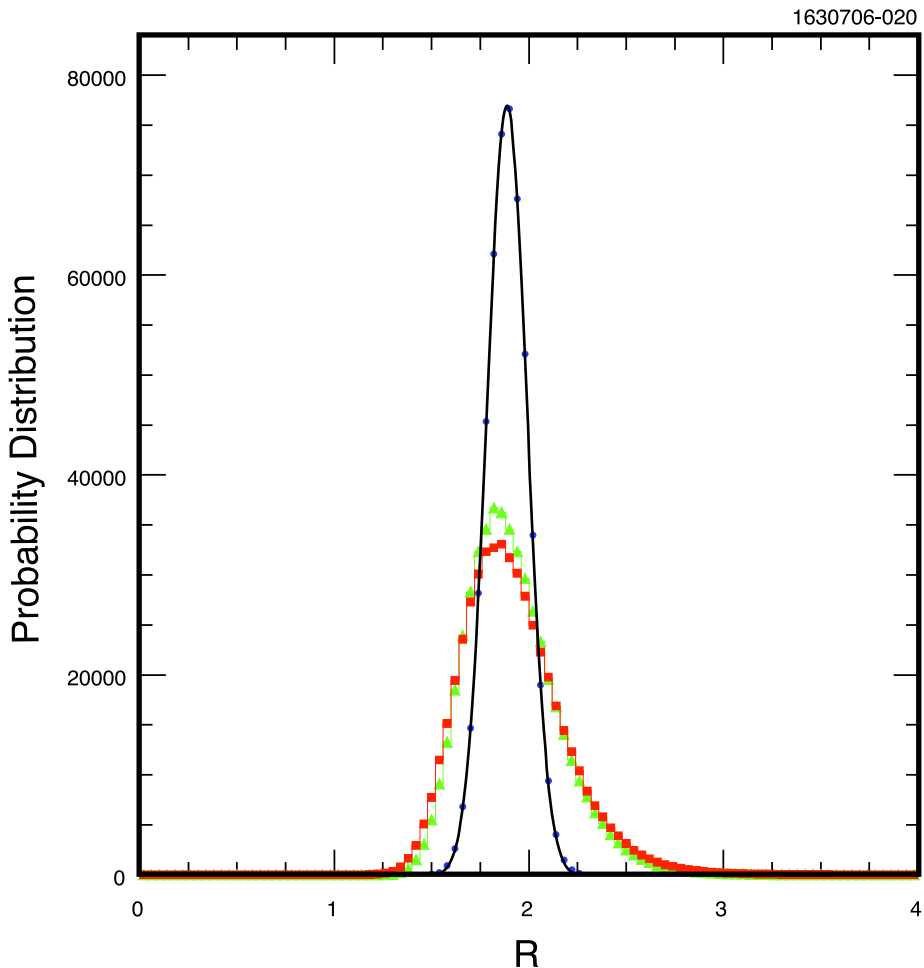,width=4.8in}}
  \caption{\label{phi_ratio_errors} Probability distributions for
  $R$ smeared by its:
  (a) statistical error (solid line),
  (b) systematic error (filled triangles) and,
  (c) total error (filled squares).}
\end{figure}

We determine the errors, including all the correlations, by using a
Monte Carlo method where we generate each independent quantity as a
Gaussian distribution using the estimated mean as the central value
and the estimated error as the width. Then we evaluate the relevant
measured quantities. Applying this method to $R$ gives the spectra
shown in Fig.~\ref{phi_ratio_errors}. We do this separately for the
statistical and systematic errors and then combine them for the
total uncertainty.
 This is the probability distribution for this ratio. It is non-Gaussian.

 We extract the statistical, systematic and total errors
on our measurements  from the values of these distributions
corresponding to $68.3\%$ (one standard deviation) of the area under
the corresponding curves (either statistical, systematic or total)
above and below the maximum position. We measure

\begin{equation}
R=1.9\pm 0.1^{+0.3}_{-0.2}~.
\end{equation}


\subsection{Measurement of the \boldmath{$\Upsilon(5S)\to
B_s^{(*)}\overline{B}_s^{(*)}$} Branching Fraction}
\label{get-fs-phi-sec}

Eq.~(\ref{eq:ratio-to-phi}) and the spectrum in Fig.~\ref{PhiBrs}
demonstrate a significant excess of $\phi$ mesons at the
$\Upsilon$(5S). From these results, we can determine the fraction of
the $\Upsilon$(5S) that decays into $B_s^{(*)}\overline{B}_s^{(*)}$,
which we denote as $f_S$, in a model dependent manner. We assume
that the $\phi$ yields at the $\Upsilon(5S)$ comes from two sources,
$B$ and $B_s$ mesons. The equation linking them is

\begin{equation}
{\cal{B}}(\Upsilon(5S)\to \phi X)/2= f_S\cdot {\cal{B}}(B_s\to \phi
X)+\frac{(1-f_S)}{2} \cdot{{\cal{B}}(\Upsilon(4S)\to \phi X)}
.~~\label{get_fs_phi}
\end{equation}
We restrict our consideration to the interval $0.05<x<0.50$. We
obtain $f_S$ via the equation \cite{correction}
\begin{eqnarray}
f_S= \frac{ {\cal{B}}^{(x>0.05)}(\Upsilon(5S)\to \phi
X)-{\cal{B}}^{(x>0.05)}(\Upsilon(4S)\to \phi X) } {
2{\cal{B}}^{(x>0.05)}(B_s \to \phi
X)-{\cal{B}}^{(x>0.05)}(\Upsilon(4S)\to \phi X)
}~.\label{get_fs_phi2}
\end{eqnarray}
The branching fractions ${\cal{B}}^{(x>0.05)}(\Upsilon(5S)\to \phi
X)$ and  ${\cal{B}}^{(x>0.05)}(\Upsilon(4S)\to \phi X)$ are given in
Eqs. \ref{4s-to-phi-9} and \ref{5s-to-phi}. It is necessary to
estimate ${\cal{B}}^{(x>0.05)}(B_s\to \phi X)$.  We first show that
most of the $\phi$'s in $b$ decay result from the decay chain $B\to
(D$ or $D_s) X$, $D{\rm ~or~} D_s\to\phi X$. The decay rates for $B
\to D X$ are tabulated by the PDG \cite{PDG}. CLEO-c has measured,
in a companion paper \cite{D_Ds_incl}, the branching ratios for
$D^0$, $D^+$ and $D_s^+$ mesons into $\phi$ mesons. These results
are listed in the second line of Table~\ref{tab:Dinc2}.
\begin{table}[ht]
\begin{center}
\caption{ $B$ decay inclusive branching ratios into $D^0$, $D^+$ and
$D_s^+$ mesons, the inclusive $D$ decay branching ratios into $\phi$
mesons and the resulting $B\to\phi X$ product branching ratios.}
\begin{tabular}{cccc}
\hline\hline
                                                                          &~ $D^0$ (\%)                 &~ $D^+$ (\%)             &~ $D_s^+$ (\%) \\
\hline
${\cal{B}}(B\to (D$ or $D_s) X)$                                          &~ $64.0 \pm 3.0$             &~ $22.8\pm1.4$           &~ $8.6\pm1.2$         \\
${\cal{B}}((D$ or $D_s) \to \phi X)$                                      &~ $1.05 \pm 0.10$   &~ $1.03\pm 0.12$ &~ $16.1\pm 1.6$ \\
${\cal B}(B \to (D$ or $D_s)X) \cdot {{\cal B}}((D$ or $D_s) \to \phi X)$ &~ $0.67 \pm 0.08$            &~ $0.23\pm 0.03$         &~ $1.4\pm0.2$ \\
\hline\hline
\end{tabular}
\label{tab:Dinc2}
\end{center}
\end{table}

The rates on the last line of Table~\ref{tab:Dinc2} give the
resulting $\phi$ yields from the decay of $B$ into each of the
individual charm mesons and the subsequent decay of the charm mesons
into a $\phi$.  The sum of these product rates is (2.3$\pm$0.3)\%.
The difference with the measured $B\to\phi X$ rate given above in
Eq.~(\ref{eq:Btophiss}) is an additional unaccounted for
(1.2$\pm$0.4)\%. This rate can be accounted for from the production
of charm baryons or merely by fragmentation processes. The sum of $B
\to D X$ and $D_s X$ branching ratios Table~\ref{tab:Dinc2} is
$(95.4\pm 3.5)$\%. We now assume that the number of $D$ plus $D_s^+$
mesons produced in $B_s$ decays is the same as in $B$ decays. Using
our previous estimate of ${\cal{B}}(B_s\to D_s X)=(92\pm 11)$\%
\cite{Incl_Bs_5s}, we have (8$\pm$11)\% of the $B_s$ rate into
charmed mesons that must be accounted for by a mixture of $D^0$ and
$D^+$, both of which have an equal decay rate into $\phi$'s. The
predicted rates are
\begin{eqnarray}
{\cal{B}}(B_s\to \phi X)&=&(16.1\pm2.4)\%\\
 {\cal{B}}^{(x>0.05)}(B_s\to\phi X)&=&(15.7\pm2.3)\%,
\end{eqnarray}
where we have added in the (1.2$\pm$0.4)\% from other processes; the
rate for $0.05>x>0$ is taken from our Monte Carlo simulation, and
amounts to 2.6\% of the total yield.

\begin{figure}[htbp]
\centerline{\epsfig{figure=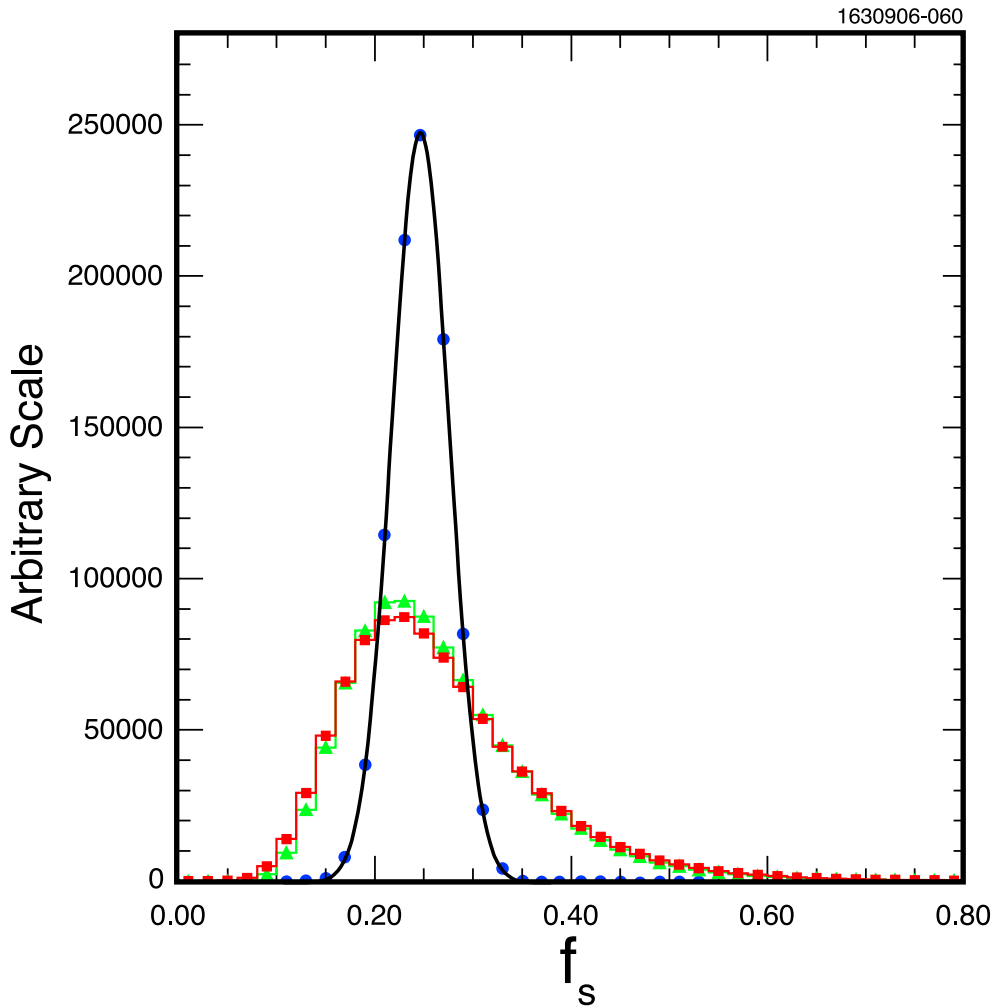,width=5.2in}}
  \caption{\label{fs-phi-errors} Probability distribution of
  ${\cal{B}}(\Upsilon(5S)\to B_s^{(*)}\overline{B}_s^{(*)})$,
  measured using the inclusive yields of $\phi$ mesons, smeared by:
  (a) statistical error (fitted dots),
  (b) systematic error (filled triangles) and,
  (c) total error (filled squares).
 }
\end{figure}

Solving Eq.~(\ref{get_fs_phi2}) and using our procedure for finding
the errors by generating Gaussian distributions for the independent
quantities leads to the probability distributions of $f_S$ shown in
Fig.~\ref{fs-phi-errors}.
We measure the $B_s^{(*)}\overline{B_s}^{(*)}$ ratio to the total
$b\overline{b}$ quark pair production above the four-flavor ($u$,
$d$, $s$ and $c$ quarks) continuum background at the $\Upsilon$(5S)
energy as

\begin{equation}
{\cal{B}}(\Upsilon(5S)\to B_s^{(*)}\overline{B}_s^{(*)})=( 24.6 \pm
 2.9 ^{+11.0}_{-~5.3} )\%~.~
\end{equation}

\subsection{Estimate of \boldmath{${\cal{B}}(\Upsilon({\rm 5S})\to \phi X)$}}

We estimate the branching fraction in the 0$<x<$0.05 interval for
$\Upsilon$(5S) decays by using the ratio of the $\phi$ meson yield
in the first $x$ interval to the total. This estimate is obtained
from a combination of $B_s\overline{B_s}X$ and $B\overline{B}X$
Monte Carlo simulated decays at the $\Upsilon$(5S) energy. To
combine these two types of events, we use the average of our two
measurements of the $f_S$ fraction, 19.3\%. Here the fraction in the
first bin from $B$ decays is 2.4\% very similar to the 2.6\% from
$B_s$ decays, that makes the error from $f_S$ negligible.  Adding
this estimate to the ${\cal{B}}^{(x>0.05)}(\Upsilon(5S)\to \phi X)=(
12.9 \pm 0.7 ^{+2.2}_{-1.4} )\%$ as measured in
Eq.~(\ref{5s-to-phi}), we find

\begin{equation}
{\cal{B}}(\Upsilon(5S)\to \phi X)=( 13.8 \pm 0.7 ^{+2.3}_{-1.5}
)\%.~\label{5s-to-phi2}
\end{equation}


\section{Updated Measurement of \boldmath{${\cal{B}}(\Upsilon(5S) \to
B_s^{(*)}\overline{B_s}^{(*)})$} Using \boldmath{$D_s$} Meson
Yields}

Here we update our measurement of $f_s$ using $D_s$ yields given in
Ref.~\cite{Incl_Bs_5s} due to changes in the scale factors
(Eqs.~\ref{eq:4s_scale_Factor1} and \ref{eq:5s_scale_Factor1}) and
therefore to a better estimate of the number of hadronic events. We
find the ratio

\begin{equation}
\frac{{\cal{B}}(\Upsilon(5S)\to D_s X)}{{\cal{B}}(\Upsilon(4S)\to
D_s X)}= 2.5 \pm 0.2 ^{+0.4}_{-0.3}~.\label{eq:Ratio}
\end{equation}

We rewrite Eq.~(\ref{get_fs_phi}) for $D_s$ mesons rather than for
$\phi$ mesons as

\begin{eqnarray}
{\cal{B}}(\Upsilon(5S)\to D_s X)\cdot{{\cal{B}}(D_s \to \phi\pi)}/2=
f_S\cdot {\cal{B}}(B_s\to D_s X)\cdot{{\cal{B}}(D_s \to
\phi\pi)}\\
+\frac{(1-f_S)}{2} \cdot{{\cal{B}}(\Upsilon(4S)\to D_s X)}
\cdot{{\cal{B}}(D_s \to \phi\pi)}~.~\label{get_fs_Ds}
\end{eqnarray}
Then using the product of production rates in Ref.~\cite{Incl_Bs_5s}
updated with the new scale factors, our model dependent estimate of
${\cal{B}}(B_s\to D_s X)=( 92 \pm 11 )\%$~\cite{Incl_Bs_5s} and the
newest estimate of ${\cal{B}}(D_s\to \phi \pi)=( 4.4 \pm 0.6
)\%$~\cite{PDG}, we obtain the $B_s^{(*)}\overline{B_s}^{(*)}$ ratio
to the total $b\overline{b}$ quark pair production above the
four-flavor ($u$, $d$, $s$ and $c$ quarks) continuum background at
the $\Upsilon$(5S) energy of

\begin{equation}
f_S={\cal{B}}(\Upsilon(5S)\to B_s^{(*)}\overline{B}_s^{(*)})=( 16.8
\pm 2.6 ^{+6.7}_{-3.4} )\%~\label{eq:5stoBsEstimated}
\end{equation}

\begin{figure}[htbp]
\centerline{\epsfig{figure=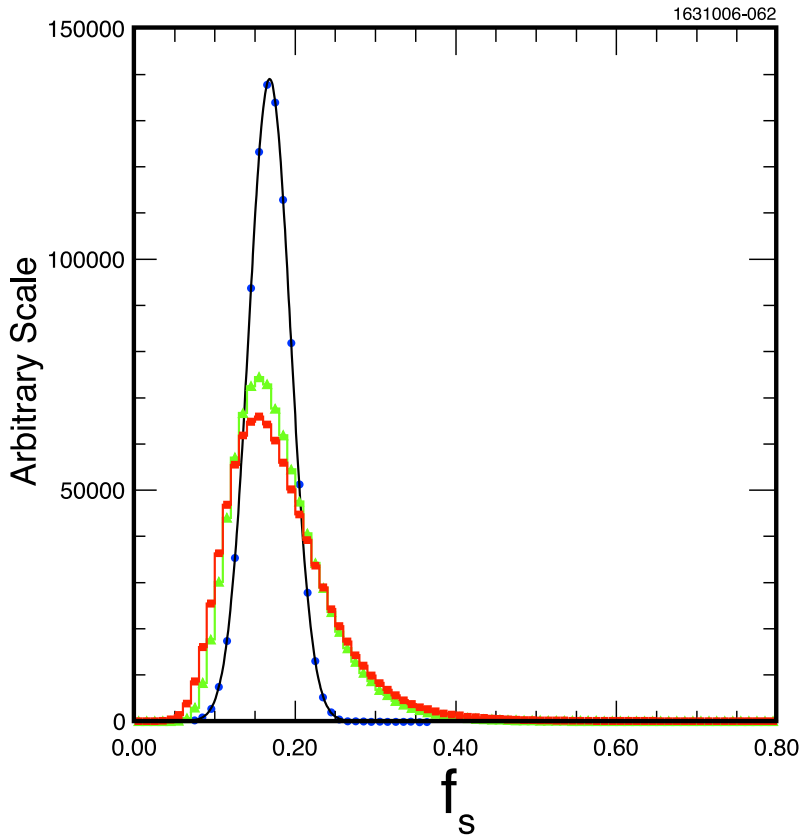,width=4.2in}}
  \caption{\label{fs-ds-errors} Probability distributions for
  ${\cal{B}}(\Upsilon(5S)\to B_s^{(*)}\overline{B}_s^{(*)})$,
  obtained by measuring $D_s$ yields, and
  smeared by:
  (a) statistical error (filled circles),
  (b) systematic error (filled triangles) and,
  (c) total error (filled squares).
  }
\end{figure}

The probability distribution of this measurement within the
statistical, the systematic and the total error is shown in
Fig.~\ref{fs-ds-errors}.

\begin{figure}[htbp]
\centerline{\epsfxsize=6.7in
\epsffile{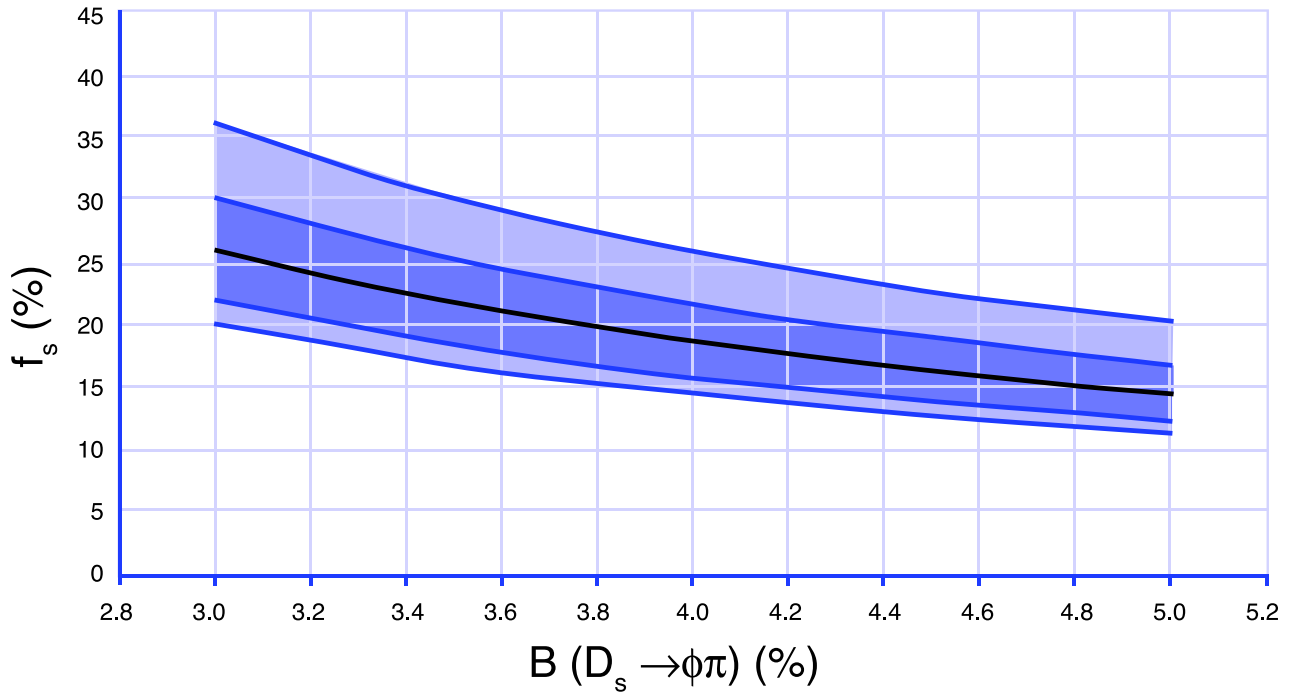}}
  \caption{\label{dependence_on_dstophipi} $f_s$ versus
  ${\cal{B}}(D_s \to \phi \pi)$ (the central line), the statistical
  error on $f_s$ is represented by the inner adjacent lines and the total error
  (the statistical and systematic added in quadrature) is represented by the outer two lines. The
  error on ${\cal{B}}(D_s \to \phi \pi)$ is not included.}
\end{figure}

This measurement is obtained using a value of $(4.4 \pm
0.6)\%$~\cite{PDG} for the $D_s \to \phi\pi$ branching fraction.
Since this branching fraction is not well known, we show in
Fig.~\ref{dependence_on_dstophipi} the behavior of $f_s$ for
different values of ${\cal{B}}(D_s \to \phi\pi)$ ranging from $3.0$
to $5.0$\%.

\section{The \boldmath{$\Upsilon(5S)$} Production Cross Section and
\boldmath{$B$} Meson Yields}

We measure the cross-section
\begin{equation}
\sigma(e^+e^- \to \Upsilon(5S))= (0.301 \pm 0.002 \pm 0.039)~{\rm
nb} ~,\label{eq:xsec}
\end{equation}
where the first error is statistical and the second is systematic
and due to the error in the relative luminosity measurement between
$\Upsilon$(5S) and continuum.

Measurement of this cross-section allows us to present the previous
CLEO exclusive decay results for $B$ mesons at the $\Upsilon$(5S)
\cite{B_5s} in terms of branching fractions. These are given in
Table~\ref{5s_decays}. These results are consistent with theoretical
expectations \cite{Models,UQM}.

The inclusive $B$ meson yield is represented as $B \overline{B}X$ in
the Table. This measurement of ($58.9\pm10.0\pm9.2)$\% is equal to
1-$f_S$ under the assumption that there are no non-$b\overline{b}$
decays of the $\Upsilon$(5S).

\begin{table}[htb]
\begin{center}
\end{center}
\caption{Branching fractions of the possible $\Upsilon$(5S) final
states either measured directly in this paper or estimated using our
measurement of the $\Upsilon$(5S) total cross section and the cross
section and upper limits reported in Ref.~\cite{B_5s}.}
\label{5s_decays}
\begin{tabular}{lcc}
\hline\hline
Different components   & Measured       &  Corresponding   \\
of the $\Upsilon$(5S)  & cross-section (nb)  &  Branching Fraction ($\%$) \\
\hline
$B_s^{(*)} \overline{B}_s^{(*)}$                      &~ $-$                              ~&~ $(16.8 \pm 2.6 ^{+6.7}_{-3.4} )\%$   \\
                     &~                               ~&~ (using $D_s$ yields)   \\
$B_s^{(*)} \overline{B}_s^{(*)}$                      &~ $-$                              ~&~ $(24.6 \pm 2.9 ^{+11.0}_{-~5.3})\%$   \\
                      &~                               ~&~ (using $\phi$ yields)   \\
$B \overline{B}X$                                     &~ $0.177\pm0.030\pm0.016$  ~&~ $58.9\pm10.0\pm9.2$   \\
$B^* \overline{B}^*$                                  &~ $0.131\pm0.025\pm0.014$  ~&~ $43.6\pm8.3\pm7.2$   \\
$B \overline{B}^*+B^* \overline{B}$                   &~ $0.043\pm0.016\pm0.006$  ~&~ $14.3\pm5.3\pm2.7$   \\
$B \overline{B}$                                      &~ $<0.038$ ~ (@ 90\% CL) ~&~ $<13.8$   \\
$B \overline{B}^{(*)}\pi+B^{(*)} \overline{B}\pi$     &~ $<0.055$ ~ (@ 90\% CL) ~&~ $<19.7$   \\
$B \overline{B}\pi\pi$                                &~ $<0.024$ ~ (@ 90\% CL) ~&~ $<8.9$   \\
\hline\hline
\end{tabular}
\end{table}

\section{Conclusions}

We measure the momentum spectra of
 $\phi$ mesons from $\Upsilon$(4S) and
$\Upsilon$(5S) decays, and use these data to estimate the ratio
$f_S$ of $B_s^{(*)}\overline{B_s}^{(*)}$ to the total
$b\overline{b}$ quark pair production at the $\Upsilon$(5S) energy
as $( 24.6 \pm 2.9 ^{+11.0}_{-~5.3} )\%$. We also update our
previous measurement of $f_S$ using $D_s$ yields
\cite{Incl_Bs_5s}, and we find $( 16.8 \pm 2.6 ^{+6.7}_{-3.4}
)\%$. The central value and the error slightly change due to a
better determination of the relative luminosity. Furthermore,
using our cross-section measurement of $\sigma_{5S}= (0.301 \pm
0.002 \pm 0.039)$ nb for hadron production above 4-flavor
continuum at the $\Upsilon$(5S) energy, we measure total $B$ meson
production and extract $f_S = (41.1\pm10.0\pm9.2$)\%. Taking a
weighted average of all three methods gives
$f_S=(21^{+6}_{-3})$\%, where common systematic errors have been
accounted for.

 Our results as summarized in Table~\ref{tab:summary}, and at the
current level of precision, are consistent with the previously
published phenomenological predictions \cite{Models,UQM}.

\begin{table}[htbp]
\begin{center}
\caption{Summary of our results.} \label{tab:summary}
\begin{tabular}{lc}
\hline\hline
Quantity & Measurement \\

\hline\multicolumn{2}{l}{Results from inclusive $\phi$
measurements}\\

${\cal{B}}(\Upsilon(4S)\to \phi X)$ & $( 7.06 \pm 0.09 \pm 0.60 )\%~$  \\
 ${\cal{B}}(B\to \phi X)$ & $( 3.53 \pm 0.05 \pm 0.30 )\%~$ \\
 ${\cal{B}}(\Upsilon(5S)\to \phi X)$ & $( 13.8 \pm 0.7 ^{+2.3}_{-1.5} )\%~$  \\
 ${{\cal{B}}(\Upsilon(5S)\to \phi X)}/{{\cal{B}}(\Upsilon(4S)\to \phi X)}$  & $1.9 \pm 0.1 ^{+0.3}_{-0.2}~$  \\
${\cal{B}}(\Upsilon(5S)\to B_s^{(*)}\overline{B}_s^{(*)})$ &
$( 24.6 \pm 2.9 ^{+11.0}_{-~5.3} )\%~$  \\
\hline \multicolumn{2}{l}{Results from inclusive $D_s$
measurements}\\

${{\cal{B}}(\Upsilon(5S)\to D_s
X)}/{{\cal{B}}(\Upsilon(4S)\to D_s X)}$ & $2.5 \pm 0.2 ^{+0.4}_{-0.3}~$  \\
${\cal{B}}(\Upsilon(5S)\to B_s^{(*)}\overline{B}_s^{(*)})$ &
$(16.8 \pm 2.6 ^{+6.7}_{-3.4} )\%~$ \\
\hline \multicolumn{2}{l}{Results from cross-section and $B$
measurements}\\
$\sigma(e^+e^- \to \Upsilon(5S))$ &  $( 0.301 \pm 0.002~ \pm 0.039 )~$nb \\
${\cal{B}}(\Upsilon(5S)\to B_s^{(*)}\overline{B}_s^{(*)})$
 &
$( 41.1 \pm 10.0\pm 9.2 )\%~$\\
\hline \multicolumn{2}{l}{$f_S$ from combining inclusive $\phi$, $D_s$ and $B$ measurements}\\
${\cal{B}}(\Upsilon(5S)\to B_s^{(*)}\overline{B}_s^{(*)})$ &
$( 21^{+6}_{-3} )\%~$ \\

\hline\hline
\end{tabular}
\end{center}
\end{table}

\section{Acknowledgements}
We gratefully acknowledge the effort of the CESR staff in providing
us with excellent luminosity and running conditions.
D.~Cronin-Hennessy and A.~Ryd thank the A.P.~Sloan Foundation. This
work was supported by the National Science Foundation, the U.S.
Department of Energy, and the Natural Sciences and Engineering
Research Council of Canada.

\afterpage{\clearpage}


\begin{thebibliography}{99}
\bibitem{correction}
We ignore $\sim$1\% corrections due to non-scaling changes of the
yields in the first $x$ bin.
\end{thebibliography}
\end{document}